\documentclass[12pt]{article}
\usepackage{authblk}
\usepackage[a4paper,bindingoffset=0.2in,left=0.6in,right=1in,top=0.85in,bottom=1in,footskip=.25in]{geometry}
\usepackage[dvipsnames]{xcolor}
\usepackage{hologo}
\usepackage[]{colortbl}
\usepackage{multirow}
\usepackage[natbibapa]{apacite}

\usepackage{graphics}
\usepackage{enumitem}
\usepackage{graphicx}

\usepackage[english]{babel}
\addto{\captionsenglish}{%
    }

\usepackage[breaklinks,colorlinks = true,
            linkcolor = BlueViolet,
            urlcolor  = BlueViolet,
            citecolor = Aquamarine,
            anchorcolor = white,
            breaklinks]{hyperref}
\usepackage[]{doi}

\def\mybibdoicolor{\color{BlueViolet}} 

\def\DEG{$^\circ$}

\title{{\bf A possible case of sporadic aurora observed at Rio de Janeiro\footnote{Paper published in {\it Earth, Planets and Space}, Special issue {\it Solar-Terrestrial Environment Prediction: Toward the Synergy of Science and Forecasting Operation of Space Weather and Space Climate}, doi: \url{https://doi.org/10.1186/s40623-020-01208-z}}}}

\date{\vspace{-6ex}}

\author[1,2\footnote{Electronic address: denny.m.deoliveira@nasa.gov; denny@umbc.edu}]{{\normalsize Denny M. Oliveira}}
\author[3,4,5]{\normalsize Hisashi Hayakawa}
\author[6,1]{\normalsize Ankush Bhaskar}
\author[2]{{\normalsize Eftyhia Zesta}}
\author[7]{\normalsize Geeta Vichare}

\affil[1]{{\footnotesize Goddard Planetary Heliophysics Institute, University of Maryland, Baltimore County, Baltimore, MD, United States}}
\affil[2]{{\footnotesize NASA Goddard Space Flight Center, Greenbelt, MD, United States}}
\affil[3]{{\footnotesize Institute for Space-Earth Environmental Research, Nagoya University, Nagoya, Japan}}
\affil[4]{{\footnotesize Institute for Advanced Researches, Nagoya University, Nagoya, 4648601, Japan}}
\affil[5]{{\footnotesize Rutherford Appleton Laboratory, Chilton, United Kingdom}}
\affil[6]{{\footnotesize Catholic University of America, Washington D.C., United States}}
\affil[7]{{\footnotesize Indian Institute of Geomagnetism, Plot 5, Sector 18, New Panvel (West), Navi Mumbai, India}}

\begin{document}

	\maketitle

    \begin{abstract}
      Being footprints of major magnetic storms and hence major solar eruptions, mid- to low-latitude aurorae have been one of the pathways to understand solar-terrestrial environments. However, it has been reported that aurorae are also occasionally observed at low latitudes under low or even quiet magnetic conditions. Such phenomena are known as ``sporadic aurorae". We report on a historical event observed by a scientist of the Brazilian Empire in Rio de Janeiro on 15 February 1875. We analyze this event on the basis of its spectroscopic observations, along with its visual structure and coloration, to suggest this event was a possible case of sporadic aurorae. Given the absence of worldwide aurora observations on that day as a consequence of low magnetic activity recorded on the days preceding the observation, in addition to a detailed description, the event observed can most likely be classified as a sporadic aurora. We discuss the geographic and magnetic conditions of that event. Thus, we add a possible case of sporadic aurora in the South American sector.
    \end{abstract}

    \section{Introduction}

      Extreme space weather events, such as the events of August-September 1859 \citep{Carrington1859,Kimball1960,Green2006,Farrona2011a,Hayakawa2016,Gonzalez-Sparza2018,Hayakawa2018b,Hayakawa2019b}, February 1872 \citep{Meldrun1872,Silverman2008,Hayakawa2018a}, and May 1921 \citep{Silverman2001, Love2019b}, cause extremely intense magnetic storms. Among other effects, one of the most interesting visual phenomena is the occurrence of very intense and bright aurorae. Aurorae during extreme events are not only observed at high latitudes, but at low latitudes (22-23\DEG) as well \citep{Kimball1960,Silverman1995,Green2006,Humble2006,Silverman2008,Cardenas2016,Hayakawa2016,Gonzalez-Sparza2018,Hayakawa2018b,Hayakawa2019b}. \par

      As such, mid- and low-latitude aurorae have formed footprints of major magnetic storms and hence major geo-effective solar eruptions \citep{VallanceJones1992,Shiokawa2005,Silverman2006,Willis2006,Willis2009,Schlegel2011}. Therefore, such auroral reports have been one of the key pathways to understand solar-terrestrial interactions in the past in terms of their long-term variability and cyclicity \citep{Silverman1992,Usoskin2013b,Usoskin2015,Lockwood2015,Lockwood2016b,Vazquez2016,DominguezCastro2016}. \par

      However, auroral phenomena have been rarely seen at low latitudes during moderate and even quiet magnetic conditions. Such events are known as sporadic aurorae \citep{Silverman2003}. \cite{Botley1963} introduced this term to the scientific community, citing previous descriptive usages of the same word by \cite{Abbe1895}. She introduced nine cases of low latitude aurorae in Europe and in the Middle East observed in the 12th and 19th centuries. \cite{Botley1963} was the first to clearly define sporadic aurorae as “comprise such instances as a single ray in a sky otherwise seemingly clear of auroral light, or isolated patches well to the equatorial side of a great display". \cite{Botley1963} also noted references to reports of two low-latitude aurora occurrences without the occurrence of high-latitude aurorae \citep{Fritz1881,Eddie1894}. \par

      It took another four decades for the next paper on sporadic aurora to be published. \cite{Silverman2003} provided a survey of considerable sporadic aurora observations in low-latitude regions of the United States during a time span of over half a century, and highlighted the occurrence of sporadic aurorae in the context of mid- to low-latitude aurorae during moderate to low magnetic activity. That paper was later followed by other papers with reports on sporadic aurora sightings from Iberia and the Canary Islands \citep{Vaquero2007a,Vazquez2010}, East Asia \citep{Willis2007}, Mexico \citep{Vaquero2013}, and the Philippines \citep{Hayakawa2018c}. Interestingly, \cite{Shiokawa2005} reported three cases of instrumental observations of mid-latitude aurora in Hokkaido (Japan) under fairly moderate magnetic activity as well. \par

      \cite{Silverman2003} speculated that sporadic aurorae may be caused by localized and ephemeral magnetospheric energy input into the low-latitude ionosphere, but he does not clearly suggest any physical mechanisms that may explain this phenomenon. In fact, considering the known correlation between intensity of magnetic disturbance and equatorward boundary of auroral ovals \citep{Yokoyama1998}, \cite{Silverman2003}'s comprehensive survey was striking and casted an open question on its physical mechanism. \cite{Hayakawa2018c} suggested that at least part of sporadic aurorae might have been caused by the impact of inclined interplanetary shocks \citep[see also][]{Oliveira2018b,Oliveira2018a} that strike the magnetosphere in the pre-dusk sector. However, despite all these efforts, a comprehensive understanding of the causes of sporadic aurorae still remains an open question in space weather research. \par

      The main goal of this article is to show an aurora observation report published in a Rio de Janeiro's newspaper on 17 February 1875, hitherto unknown to the scientific community. Based on the event descriptions, the expertise of the observer, and the sporadic aurora characteristics presented in this introduction, as well as the magnetic latitude location of Rio de Janeiro and the low magnetic activity on the days before the observation, we will show that the event was most likely a sporadic aurora. The paper is structured as follows. Section 2 brings brief descriptions of the observational site and the observer. Section 3 introduces the report along with its interpretation based on current aurora knowledge. Finally, the paper is concluded in section 4 along with a final remark.

      \section{The observational site and the observer}

        \subsection{The Imperial Observatory}

          Brazil was a Portuguese colony during the period 1500 to 1822. Due to military and commercial sanctions imposed by Napoleon to Lisbon in the beginning of the 19th century, the throne of the Portuguese Empire exiled from Lisbon to Rio de Janeiro in 1808 \citep{Fausto1994}. Later, John VI of Portugal returned back to Lisbon and left his son Peter I as the ruler of the Kingdom of Brazil. Then, on 7 September 1822, Peter I proclaimed Brazil's independence of Portugal, and became the first emperor of Brazil \citep{Fausto1994}. In 1827, seven years before his death, Peter I founded the Imperial Observatory \citep{Morize1987}, today known as the National Observatory ({\it Observat\'orio Nacional}), still located in Rio de Janeiro. After Peter I's death, his son Peter II became the second and last emperor of Brazil, when it became a Republic on 15 November 1889 \citep{Fausto1994}. Peter II was a monarch very interested in science who supported many contemporary scientists, and used the auspices of the Imperial Observatory for astronomical observations and scientific discussions \citep{Benevides1979}.

        \subsection{Emmanuel Liais}

          The likely sporadic aurora reported here was observed from the facilities of the Imperial Observatory by the Frenchman Emanuel Liais (1828-1900) on 15 February 1875. Liais was the director of the Imperial Observatory in 1875, having been directly appointed by Peter II, after leaving the position as the director-adjunct of the Paris Observatory in France \citep{Morize1987}. The observer was a professional 19th century scientist. While at the Imperial Observatory, Liais conducted research on astronomy with emphasis on planetary motion and comets, discovering one himself in 1860 \citep{Liais1860}. He also published a popular book on astronomy \citep{Liais1865}. Liais had considerable experience and expertise with optical physics and instrumentation. He published on the 1858 total solar eclipse observation from Brazil, being among the first to photograph the solar corona \citep{Liais1861,Aubin2016}, and apparently had a good understanding of atmospheric effects with respect to their altitude occurrences \citep{Liais1859,MunizBarreto1997}. Liais also published on aurora observations from his home town Cherburg, France, on the Halloween day of 1853 \citep{Liais1853}. More surprisingly, Liais even suggested methods to measure auroral altitudes, showing that aurorae occur far higher than meteorological phenomena \citep{Liais1851}, as is well known today \citep[e.g.,][]{Roach1960}. According to \cite{MunizBarreto1997}, these findings would have contributed to classify auroral phenomena as magnetic phenomena as opposed to meteorological phenomena if they had been published in a scientific journal with higher audience.

      \section{The report and its interpretation}

        \subsection{Presentation of the report}

          Emmanuel Liais observed the aurora event on 15 February 1875 from Rio de Janeiro, Brazil. At that time, the Imperial Observatory was hosted by the {\it Morro do Castelo} (Castle Hill), an old church whose geographic coordinates are 22.75$^\circ$S, 43.10$^\circ$W. Liais took notes of his observations and wrote a report to the local \cite{JornalDoCommercio1875} (Commerce Newspaper). This report was found in the data base of the National Digital Library of the National Library of Brazil (\url{http://bndigital.bn.gov.br}), hereafter BNDigital). We transcribed the full text of early modern Portuguese with its original spelling and grammar style in Appendix \ref{original} and translated it into English in Appendix \ref{translation}. \par

          \begin{figure*}
            \centering
            \includegraphics[width=0.62\textwidth]{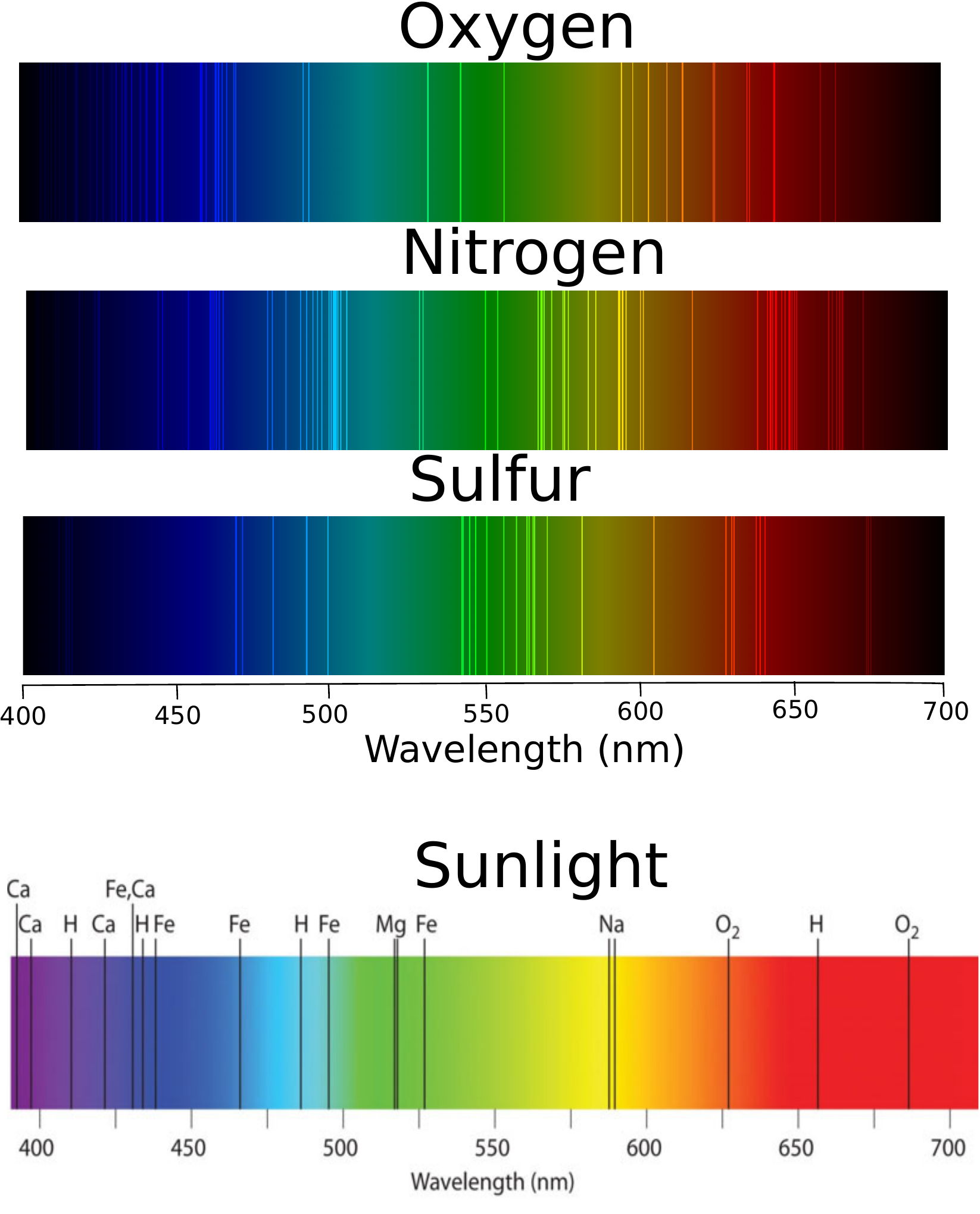}
            \caption{\footnotesize{From top to bottom, line spectra of oxygen, nitrogen, sulfur, and solar (reflected) light, with wavelengths in the horizontal axes. Whereas the auroral lights by oxygen and nitrogen show bright emission lines, the sunlight and its reflection (solar reflected light) show dark absorption lines.}}
            \label{spectrum}
          \end{figure*}

          The observer noted the occurrence of the aurora by 19:45 local mean time (LMT), or $\sim$ 16:45 GMT (Greenwich Mean Time). He realized the presence of white light in the sky that was aligned with the terrestrial magnetic field. Later, he noted that the light rays, with variable intensities, moved from west to east and took some reddish color at the bottom and greenish color at the top. This display lasted for approximately 40 minutes. \par

          While we certainly need to be careful for possible misinterpretation of atmospheric optics as aurorae \citep[e.g.,][]{Usoskin2017a,Stephenson2019}, there are three more descriptions that strongly suggest that the phenomenon observed by Liais was a sporadic aurora. First, he mentions that some clouds later passed below the auroral rays. Since Liais knew that the aurora is formed above meteorological phenomena occurring in the troposphere \citep{Liais1859,MunizBarreto1997} based on his previous experience with aurora observations \citep{Liais1853,Liais1859,Liais1865}, here it seems that he was very convinced the lights he observed were indeed auroral lights rather than meteorological phenomena associated with clouds. Secondly, its reported direction and coloration seem to rule out this kind of possible contamination. This phenomenon appeared from the direction of the magnetic needle inclination and generated white stripes in a meridian direction. This is consistent and typical with auroral ray structure, extending along the magnetic field line \citep[e.g.,][]{Chamberlain1961}. This phenomenon shows reddish color and faint greenish color well after sunset (18:40 LMT), whereas the nighttime atmospheric optics caused by the Moon is too faint to obtain its color detected by human eyes \citep{Minnaert1993}. \par 

          Even more decisively, Liais saw this phenomenon with his spectroscope and confirmed ``the certain evidence of proper lights”. Spectroscopic observations frequently give us incontrovertible evidence to distinguish aurora from other atmospheric optics, as auroral spectra show emission lines, whereas spectra of solar reflected lights show absorption lines \citep[see Figure \ref{spectrum}; e.g.,][]{Capron1879,Capron1883,Love2018,Stephenson2019}. Therefore, we can incontrovertibly reject possible contamination of atmospheric optics or clouds with strange color. \par

          Strangely, Liais interpreted the observed spectra as those of sulfur. This contradicts the modern understanding of auroral spectra as the present day understanding of the aurora shows that there are emissions mainly from oxygen and nitrogen \citep[e.g.,][]{Gault1981,Chamberlain1961}. However, it is not Liais’ originality to associate auroral spectra with sulfur. Back in mid 18th century, \cite{vanMusschenbroek1762} suggested aurorae were partially caused by burning sulfur. Liais performed his observations with a spectroscope only 6 years after the earliest spectroscopic observations of the aurora \citep{Angstrom1869}. Even in the 1870's, \cite{Capron1879} acknowledged this early hypothesis of “sulfurous vapors issuing from the earth” as a cause of the aurora, while Capron himself did not seem to agree with this supposed cause. Moreover, this misinterpretation may be justified by the similarity of spectra of sulfur emissions with spectra of mixed emissions of oxygen and nitrogen, as shown in Figure \ref{spectrum}. Since spectrum lines are considered a ``fingerprint" of a source or object, this is a very important observation to distinguish aurorae form other optical phenomena. Therefore, most likely Liais was influenced by such early scientific discussions and misinterpreted the compound spectra of excited nitrogen and oxygen emissions as the emission spectrum of sulfur, as is understandable from the comparison of their spectra in Figure \ref{spectrum}. \par

          Unlike aurorae, atmospheric optics or clouds cannot shine by themselves. The light source for clouds is sunshine or solar reflected light, including moonlight \citep[e.g.,][]{Capron1883,Love2018,Stephenson2019}. Therefore, as opposed to auroral emissions, the spectra of such atmospheric optics must inevitably involve dark absorption lines, typical with the sunlight \citep[see Figure \ref{spectrum}; e.g.,][]{Capron1879}. As Liais saw this phenomenon with a spectroscope and associated it with aurora, we cannot associate this phenomenon with atmospheric optics, originated from the sunshine or solar reflected lights.

        \subsection{Modern interpretation of the report}

          There are only a few magnetic field observations that were regularly recorded around the world during the 19th century. The only magnetic indices that can be used for that period are the ak index \citep{Nevanlinna2004} and the aa index \citep{Mayaud1972}. Unfortunately, there is no ak index for that date, but there is aa index for that date. The aa index is a 3-hour time resolution magnetic index derived from two magnetic observatories in England and Australia that are nearly antipodal to each other \citep{Mayaud1972}. The Aa index is then derived from the aa index by taking its daily averages. More detail of these indices can be found in the literature \citep{Rostoker1972,Mayaud1980a}. The aa and Aa indices are provided by the British Geological Survey website. Magnetic latitudes are computed by the geomagnetic field GUFM1 model \citep{Jackson2000} from 1600 to 1990. This model is complimentary to the International Geomagnetic Reference Field (IGRF) model \citep{Thebault2015}. IGRF can compute magnetic fields from 1900 onwards, but GUFM1 can compute magnetic fields as far back as 1590 due to the compilation of a massive data base obtained from observational logs compiled on ships at sea and ports around the world \citep{Jackson2000,Jonkers2003}. \par

          \begin{figure}
            \centering
            \includegraphics[width=0.90\textwidth]{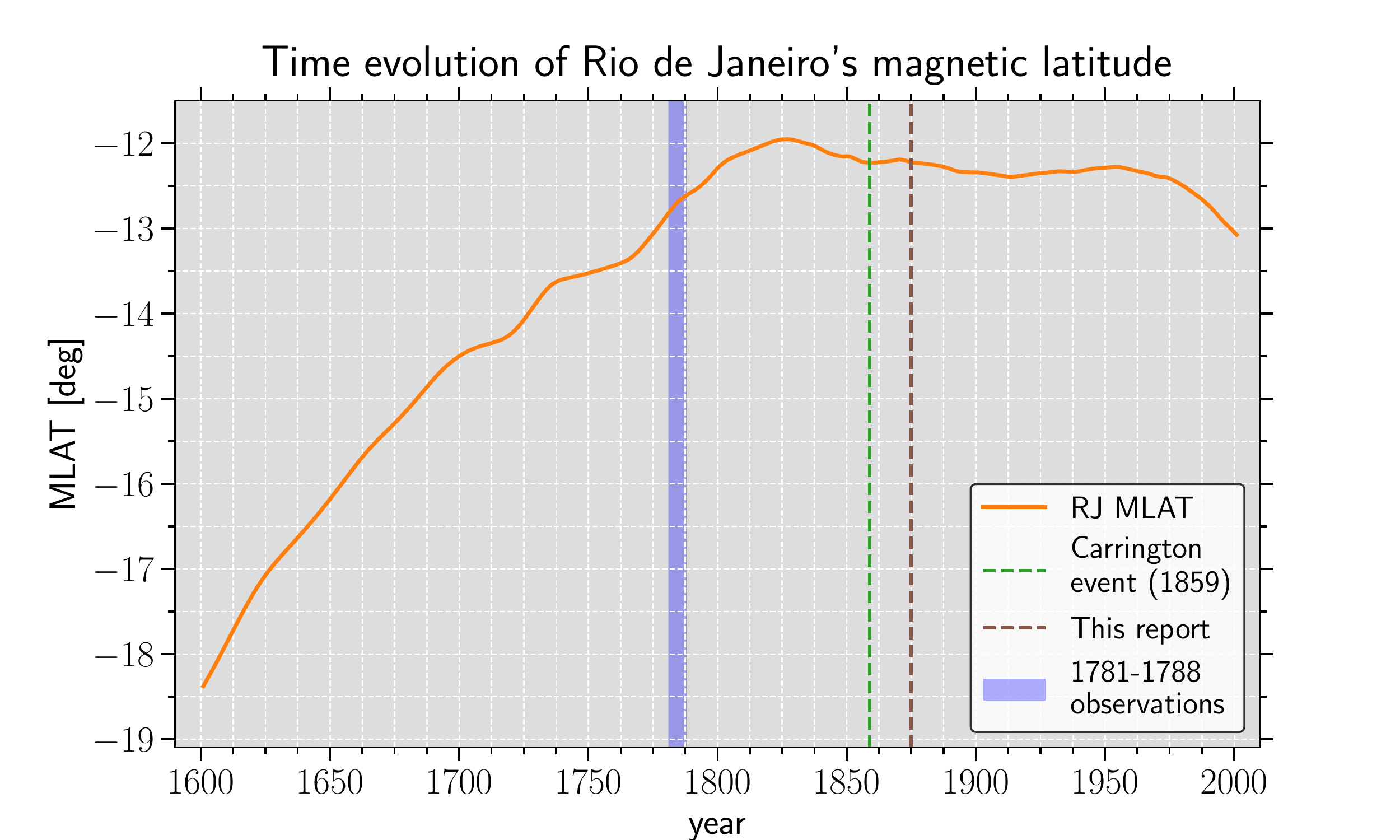}
            \caption{\footnotesize{Time evolution of Rio de Janeiro's magnetic latitude computed with the geomagnetic field model GUFM1 \citep{Jackson2000}. The light purple bar shows the period corresponding to Sanches Dorta's magnetic and aurora observations in Rio de Janeiro during the period 1781-1788 \citep{Vaquero2005,Vaquero2006}. The green and brown vertical lines mark the Carrington event \citep[e.g.,][] {Green2006,Hayakawa2018b,Hayakawa2019b} and the eventual sporadic aurora observation here reported.}}
            \label{rj_mlat}
          \end{figure}

          The solid orange line in Figure \ref{rj_mlat} shows the time evolution of Rio de Janeiro's magnetic latitude (MLAT) from 1600 to 1990. The model shows that MLAT increased from --18.4$^\circ$ in 1600 to its maximum value (the closest value to the magnetic equator) slightly above --12$^\circ$ around 1816 when it started to decrease again. The highlighted light purple area (discussed later) corresponds to the 1781-1788 interval between aurorae observed from Rio de Janeiro. The dashed green vertical line marks the Carrington event occurrence (1859), while the dashed brown vertical line indicates the event reported in this letter (1875). \par

\begin{figure}
            \centering
            \includegraphics[width=0.81\textwidth]{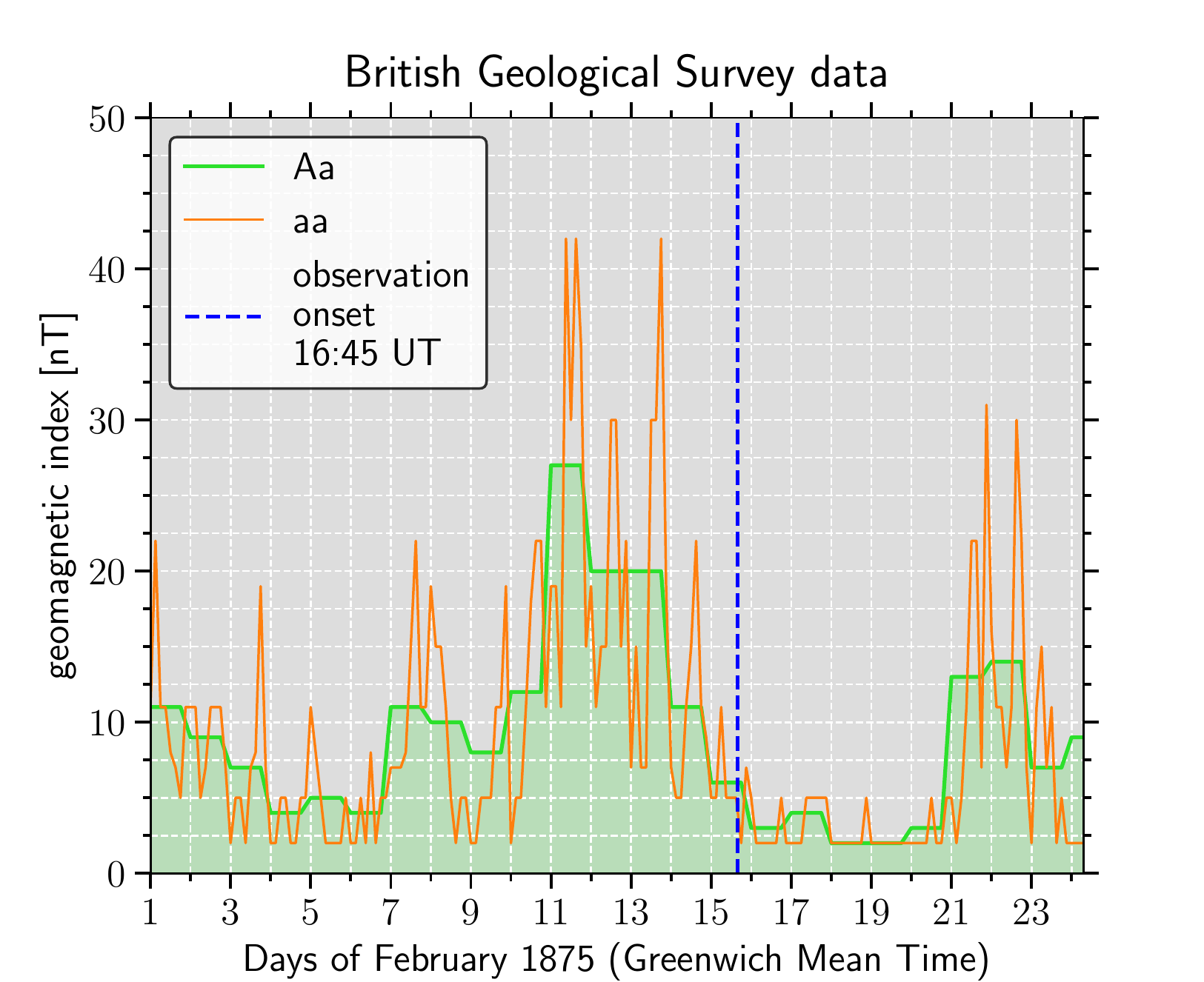}
            \caption{\footnotesize{British Geological Survey 3-hour aa index (solid orange line) and the 24-hour Aa index (solid green line/shaded green area) for the interval 1-24 February 1875. The dashed blue line indicates the onset of the aurora observation (15 February 1875 at 19:45 LMT/16:45 GMT) reported by Emmanuel Liais to the \cite{JornalDoCommercio1875}.}}
            \label{bgs}
          \end{figure}

          Figure \ref{bgs} shows the aa index for the interval 1-24 February 1875. The solid orange line indicates 3-hour aa index (in nT), while the shaded green line indicates the daily-averaged Aa index. The dashed blue line corresponds to the beginning of Liais' observations (19:45 LMT or 1645 UT) reported to the \cite{JornalDoCommercio1875}. The plot documents that the A(a)a indices showed some weak/mild activity 2-4 days prior to the aurora observation, with maximum Aa around 27 nT. This magnetic activity is consistent with sunspot number observations recorded a few days before, with very low values and one day with the observation number of 60 \citep{Clette2014,Clette2016b}. The low magnetic activity conditions during that sporadic aurora event is consistent with the description suggested by \cite{Silverman2003}. \par

          Additionally, the results of this study may also explain the reason why great aurora displays observed from Brazil have not been found/reported in the contemporary records for the Carrington event yet. As seen in Figure \ref{bgs}, Rio de Janeiro's MLAT by 1859 was very low, around --12.2$^\circ$. While we surveyed auroral reports in Brazilian newspapers during the Carrington event in the BNDigital database, we found only references to great aurora displays and even telegraph system failures in North America and Europe, with nothing being reported as having been observed from Brazil. \par

          Since the equatorward boundary of the auroral oval is reconstructed $\sim$ 28.5-30.4$^\circ$ MLAT \citep{Hayakawa2018b} assuming aurora height $\sim$ 400 km \citep{Roach1960,Ebihara2017}, the expected elevation of auroral visibility would be at best 3$^\circ$ above the horizon and hence it was quite difficult to observe auroral displays in the sky of Rio de Janeiro (and other Brazilian locations) during the Carrington event. However, it would be worth surveying potential auroral reports in Argentine, Chile, and Uruguay, countries that are located in regions of higher MLATs in South America. Another significant event, the magnetic storm of 4 February 1872, also triggered great aurora displays at low latitudes \citep{Silverman2008,Hayakawa2018a}. \cite{Silverman2008} reported on possible aurora sightings in latitudes as low as 10o or even 3\DEG, while he casted a caveat on their reliability. The author reported aurora sightings on the French Reunion Island, in the Indian Ocean (21.12\DEG{}S, 55.54\DEG{}E). We found mention to these aurora sightings during that storm in Brazilian newspapers while searching the BNDigital database, but none occurring from Rio de Janeiro or anywhere else in Brazil. \par

          Another possibility is to interpret Liais’ optical observations as equatorial plasma bubbles (EPBs), which are structures with depleted plasma density usually formed after sunset in the bottomside ionosphere and move from west to east \citep{Mendillo1983,Kelley2009,Liu2017b}. Since plasma bubbles are faint structures that can be hardly seen by the naked eye \citep[e.g.,][]{Wiens2006}, Liais’ event most likely cannot be classified as an EPB event. \par

          It should be mentioned that this is not the first report of an aurora observation performed from Rio de Janeiro, despite its proximity to the magnetic equator. In fact, \cite{Vaquero2005} and \cite{Vaquero2006} and \cite{Carrasco2017} presented a series of magnetic observations and aurora sightings conducted by the Portuguese astronomer Sanches Dorta in the 18th century. According to the authors, the observations conducted by Sanches Dorta, during the period 1781 to 1788, must very likely have occurred during times of elevated magnetic activity. However, according to Figure \ref{rj_mlat}, the MLATs of Rio de Janeiro during these events were around --12.5\DEG{} (highlighted purple area). If the events reported by \cite{Vaquero2005}, \cite{Vaquero2006} and \cite{Carrasco2017} were in fact caused by great magnetic storms, their visibility would have reached MLATs closer to the magnetic equator in comparison to the low-latitude Carrington aurorae previously reported \citep{Green2006,Hayakawa2018b,Hayakawa2019a,Hayakawa2019b}. 

      \section{Conclusions and a final remark}

        Sporadic aurorae occur at low-latitude areas \citep{Abbe1895,Botley1963} during moderate-to-low magnetic or even quiet conditions \citep{Silverman2003}. However, despite being impressive, this space weather phenomenon is not very well known by the community. In addition, this phenomenon does not happen very often, and there are only a few publications reporting on sporadic aurora sightings \citep{Abbe1895,Boyer1898,Botley1963,Silverman2003,Vaquero2007a,Vazquez2010,Vaquero2013,Willis2007,Hayakawa2018c}. \par

        In this letter, we presented for the first time a report on a possible sporadic aurora observation performed from Rio de Janeiro, Brazil, on 15 February 1875. This is the first sporadic aurora report in South America, and the second one in the southern hemisphere (the first observation was reported by \cite{Eddie1894}. Additionally, this is the second sporadic aurora observed near the magnetic equator. The original report was authored by Emmanuel Liais, then director of the Imperial Observatory of Rio de Janeiro, and published in the \cite{JornalDoCommercio1875} of the same city. Given the scientific expertise, the contents of scientific descriptions and the experience of the observer, particularly with respect to the use of a spectroscope, Liais' report may be considered credible and possible misinterpretation of the observed phenomenon, such as caused by atmospheric optics \citep{Usoskin2017a, Hayakawa2018c}, may be discarded. The aurora description presented by Liais is consistent with sporadic aurorae \citep{Abbe1895,Botley1963,Silverman2003}. In addition, the very low magnetic latitude of Rio de Janeiro and the weak/mild magnetic activity during the observations are consistent with a previous sporadic aurora observation near the magnetic equator \citep{Hayakawa2018c}. \par

        Furthermore, in addition to the sporadic aurora causes presented in the introductory section, we speculate that sporadic aurorae may also be caused by the flow of solar wind phase fronts with some inclination in the equatorial plane toward the dusk flank. As suggested by \cite{Cameron2019a}, such flows of solar wind phase fronts during times of low magnetic activity or quiet conditions would increase magnetic activity over time due to shear and viscosity effects, and the sudden release of this energy may cause sporadic aurorae. More observations and possibly numerical simulations are needed in order to test these hypotheses and advance the knowledge of sporadic aurora triggering. 

      \appendix

      \section{The report}\label{report}

        \subsection{Original transcript written in old Portuguese}\label{original}

          Rio de Janeiro, 17 de fevereiro de 1875 \\

          Jornal do Commercio \\

          {\bf Aurora austral}: O Sr. Emmanuel Liais, director do observatorio astronomico do Rio de Janeiro, enviou-nos ontem as sequintes observações que fez sobre a aurora austral, de que j\'a démos resumida not\'icia: \par

          A's 7 3/4 da noite foi a minha atenção despertada por uma especie de véo espalhado sobre todo o céo formando uma serie de listras esbranquiçadas, que começavam ao sul sobre um arco de circulo, cujo centro achava-se abaixo do horisonte, na direção da agulha magnetica de inclinação. As listras ou raios eram de tal extensão que atravessavam o céo do sul ao norte, onde convergiam para o ponto diametralmente oposto. \par

          Esta disposição, reproduzindo a forma das auroras boreaes, fez-me desde logo suppôr que bem podia ser uma aurora austral o phenomeno que presenciava; infelizmente não podia affirmal-o por causa da presença da lua momentos depois, porém, fiquei inteiramente convencido disso, graças a outras circumstancias que o acompanháram. \par

          Com effeito, depois de cinco minutos de observação, passaram de oéste para léste, e duas vezes manifestaram-se variações successivas de intensidade nos raios, como se dá frequentemente nas auroras boreaes e austraes. Além disso, passados mais alguns minutos os raios, cujas intensidades haviam augmentado, tomaram na parte inferior uma tenue côr avermelhada e na superior verde desmaiada, que não podia ser effeito da luz reflectida da lua. \par

          Observei então com o espectroscopio, onde appareciam linhas brilhantes, indicio certo de luzes proprias. Todas elas pertenciam ao enxofre, substancia que, como é sabido, encontra-se em quantidade apreciavel na atmosphera. \par

          Em seguida olhei para o norte, onde vi dous relampagos, e notei que se formavam pequenas nuvens de uma fórma variavel pelo effeito de condensação e da dissolução de vapores. Muitas nuvens caminhavam na direção de léste, um pouco ao sul, passando abaixo dos raios da aurora; ao mesmo tempo estes diminuiam de intensidade e as côres da parte inferior tinham desapparecido. \par

          Observei mais um halo fraco em volta da lua, dentro de um tenue véo de vapor, e como já o assignalou por Bravais, este halo era um tanto mais forte nas intercepções com os raios da aurora. Pouco depois estes raios começaram a encurtar e a retirar-se para o Sul. \par

          Foi então que deixando a observação ás 8 horas e 20 minutos, mandei para o Jornal a noticia deste phenomeno. Quando tornei a subir ao terraço os numerosos raios existiam ainda, porem mais curtos e mais fracos. Pelas 8 horas e 40 minutos começaram a desapparecer, e ás 9 horas só se vião pequenos vestigios delles junto ao horisonte, de léste a oeste particularmente. \par

          A's 10 horas parecia quererem formar-se outra vez dous ou tres raios, mas desappareceram pouco depois, e pequenas nuvens condensaram-se sobre diversos pontos do céo. Nada mais aconteceu até às 3 horas da madrugada, occasião em que fui chamado para vêr dous raios brilhantes que tinham reapparecido a leste, e que por causa da ausencia da luz lunar, chammavam mais a attenção. \par

          Depois de diminuirem, reapparecêram quatro raios na mesma região, mais fracos, porém, do que os primeiros, e duraram até que a luz do dia nascente veio fazer cessar de todo o phenomeno, e ao amanhecer o céo mostrou-se coberto de tenues cirrus. \par

          São estes os pormenores da observação, cujas deducções farão o objeto de uma memoria especial.

        \subsection{Translation of the original transcript into modern English}\label{translation}

          Rio de Janeiro, February 17, 1875 \\

         Commerce Newspaper \\

        {\bf Aurora australis} – Mr. Emmanuel Liais, director of the astronomical observatory of Rio de Janeiro, sent us yesterday the following observations that he made on the aurora australis, on which we already reported: \par

        At 7:45pm my attention was caught by a kind of bridal veil spread all over the sky forming a series of white stripes, that started in the south on a circular arc whose center was below the horizon, in the direction of the magnetic needle inclination. The stripes or rays were of such an extension that they passed through the sky from south to north, where they converged at the diametrically opposite point. \par

        Such disposition, producing the form of aurora borealis, made me suppose at once that it could be an aurora australis the phenomenon I witnessed; unfortunately I could not affirm that because of the Moon's presence: moments later, however, I was entirely convinced that was the case, thanks to other circumstances that accompanied it. \par

        In fact, after five minutes of observation, they passed from west to east, and twice successive appearances with a variety of ray intensities occurred, as often happens with aurora borealis and australis. In addition, after a few more minutes the rays, whose intensity had augmented, took in their inferior part a reddish color and in their superior part a faint greenish color that could not result from the effect of the light reflected by the Moon. \par

        I observed then with a spectroscope, where bright lights appeared, the certain evidence of proper lights. All lights belonged to sulfur, a substance that, as is well known, is found in large amounts in the atmosphere. \par

        Then I looked towards the north, where I saw two light bolts, and noticed that small clouds formed in a variable form by effects of condensation and dissolution of vapors. Such clouds moved towards the east direction, slightly to the south, passing below the aurora rays; at the same time the rays decreased in intensity and the colors of the inferior part had disappeared. \par

        I observed one more weak halo around the Moon inside a thin veil of vapor, and as pointed out by Bravais, such halo was somewhat stronger in the interceptions with the aurora rays. Later such rays started to get shorter and move southward. \par

        Then when I left the observation at 8:20pm I sent to the Newspaper the news of this phenomenon. When I went back up to the terrace the numerous rays still existed, however they were shorter and weaker. Around 8:40pm they began to disappear, and at 9:00pm it was possible to see only their vestiges together in the horizon, particularly from east to west. \par

        At 10:00pm two or three rays appeared to be formed once more, but they disappeared later, and small clouds condensed over several points in the sky. Nothing more took place until 3:00am, occasion on which I was called to see two very bright rays that had reappeared towards the east, which caught more attention because of the absence of lunar light. \par

        After diminishing, four rays appeared in the same region, however, weaker in comparison to the first ones, and lasted until the light of the breaking day brought the phenomenon to an end, and at dawn it was shown that the sky was covered by a faint cirrus. \par

        These are the details of the observations, whose deductions will make an object of a special memory. 

    \section*{Declarations}

      \subsection*{List of abbreviations}

        MLAT: magnetic latitude; LMT: local mean time; GMT: Greenwich mean time; BNDigital: Digital Library of the National Library of Brazil; IGRF: International Geomagnetic Reference Field; EPB: equatorial plasma bubble.

      \subsection*{Ethics approval and consent to participate}

        Not applicable.

      \subsection*{Authors’ contributions}

        DMO, who is proficient in Portuguese, surveyed the BNDigital database to search for historical accounts of auroral observations from Brazil. He also used GUFM1 to compute magnetic fields and coordinates. HH provided background of space weather events in history particularly with respect to sporadic aurorae. EZ contributed with the interpretation of the historical data and observations in the light of current auroral scientific understanding. AB contributed with interpreting spectroscopic observations. GV provided fundamental information on the ionospheric and magnetic field variations at low latitudes. All authors read and approved the final manuscript.

      \subsection*{Consent for publication}

        Not applicable.

      \subsection*{Competing interests}

        The authors declare that they have no competing interests.

      \subsection*{Availability of data and materials}

        The aa and Aa indices are available for download at the British Geological Survey website \url{http://www.geomag.bgs.ac.uk/data_service/data/magnetic_indices/aaindex.html}. Figure \ref{spectrum} and its data can be retrieved from \url{https://www.mathworks.com/matlabcentral/fileexchange/27796-spectra-v1-0?s_tid=mwa_osa_a}.

      \subsection*{Funding}

        DMO thanks the financial support of the NASA grants 13-SRITM132-0011 and HSR‐MAG142-0062, under contract with UMBC. HH acknowledges the JSPS Grand-in-Aid grant JP17J06954, JP15H05816, JP15H05812, and JP15K21709. AB acknowledges the support by the NASA Living With a Star Jack Eddy Postdoctoral Fellowship Program, administered by the Catholic University of America.

      \subsection*{Acknowledgments}

        The authors thank the National Library of Brazil (Biblioteca Nacional do Brasil) for providing and keeping a public newspaper archive (\url{www.bn.br}). The authors also acknowledge the British Geological Survey for providing the magnetic index data used in this investigation and Yusuke Ebihara for his helpful comments. Finally, we thank Martin Rehfeld for providing and making a FORTRAN code version of the GUFM1 model public at his web-service hosting GitHub website (\url{https://github.com/martinrehfeld}).

    \setlength{\bibsep}{4pt}


\end{document}